# Electrostatic Disturbances of Aerosol Atmospheric Plasma: Beaded Lightning

N.I. Izhovkina[1], S.N. Artekha[2,*], N.S. Erokhin[2], L.A. Mikhailovskaya[2]

[1]*Pushkov Institute of Terrestrial Magnetism, Ionosphere and*

*Radio Wave Propagation of RAS, Troitsk, Russia*

e-mail: izhovn@izmiran.ru

[2]*Space Research Institute of RAS, Moscow, Russia*

[*]*Corresponding author:* ORCID: 0000-0002-4714-4277, e-mail: sergey.arteha@gmail.com

*Abstract* – Numerous sources produce the ionization impact on the planetary atmosphere. The tropospheric cloudiness therewith is originated at such altitudes, which coincide with places of the maximum of the atmospheric ionization that caused by space radiation. Components of cosmic radiation, penetrating down to the stratospheric and tropospheric altitudes, produce in the first place the ionization of aerosols that brings to subsequent amplifying the activity of atmospheric vortices. The ionized aerosol constituent is of primary importance both in the initiation of plasma vortices and in the concentration of energy-mass by air vortices due to the humidity condensation. The ionization process is cascading, that causes the non-linear impact of penetrating rays, which amplifying with growing the air pollution. Various heterogeneities inside plasma subsystems provoke the random intensification of aperiodic electrostatic disturbances, which can exert a significant action on the development of vortices. Sometimes beaded lightning arises during thunderstorm. Using the approach of the Boltzmann kinetic equation, the characteristics of electrostatic oscillations for non-uniform plasma, are analytically derived. The analytical expressions are obtained for non-magnetized plasma. In the limits of the Earth's atmosphere, these solutions are strictly applicable to the origination of electrostatic modulations along the lines of force of the geomagnetic field and approximately applicable in cases, when the impact of the Earth's magnetic field is negligible. The explicit expressions are deduced both in the hot plasma approximation and in the cold plasma approximation. It is demonstrated that the formation of chain elements of discharge in hot inhomogeneous plasma is defined by the nonmonotonic distribution of charges and fields. At such conditions, the irregularities breaks up into a cellular structure.

*1. Introduction*

It is well known that various heat and ionization sources permanently impact on the atmosphere of our planet. In consequence of this, there exist many ways for charging atmospheric components, for instance, by photons, proton beams, electron shower, or by other particles, which originate with destroying of nuclei in interaction with space radiation. The solar wind yields into the magnetosphere and subsequently into the atmosphere with some average power $10^{23}$ CGSE/day. In addition, galactic cosmic rays from interstellar space bring into the solar system and subsequently into the earthly atmosphere high-energetic particles (primarily protons) in the range from 100 MeV up to 100 GeV. The high-energetic cosmic rays collide with the nuclei of atmospheric components and produce radionuclides. Protons are prevalent (~85%) in cosmic rays. Also alpha-particles (about 10%), beta-particles (about 1%) and other atomic nuclei with the serial number up to $N = 30$ are presented. The process of atmospheric ionization by any shower is cascading: each high-energetic particle ionizes approximately a million atmospheric components. Note that tropospheric cloudiness is developed at such altitudes, which coincide with the altitudes of the maximal ionization of atmospheric components via cosmic radiation. Cosmic-ray showers are believed to directly participate in manifestations of electrical phenomena during a thunderstorm. Solar-terrestrial connections possess nonlinear character that can be traced in the development of tropospheric cloudiness and in climate and weather variations (Pudovkin, Raspopov 1992; Zherebtsov et al. 2005, 2017; Zelenyi, Veselovsky 2008; Krivolutsky, Repnev 2012; Nagovitsyn 2013; Izhovkina et al. 2016; Izhovkina et al. 2018, 2019). As a rule, flares equal or more than 3 points on the Sun, can cause the origination of cosmic rays stream. The solar and global meteorological data evidence about some correlation: there synchronously arise cloudiness density grows that follow powerful flares on the Sun. It was noticed that the atmospheric transparency varies under the influence of injected cosmic radiation. So, for instance, the solar proton events leaded to the subsequent significant accelerating process of the cyclones origin and development near the southeastern coast of Greenland (Veretenenko, Tejll 2008). Many phenomena manifest with injection of the intense stream of the solar wind into the atmosphere of our planet. So, for example, the instability of the troposphere grows; the intensity of atmospheric circulation varies, and the character of distribution of surface pressure alters significantly. Therefore, the totality of similar features for such phenomena evidences the possible trigger mechanism of their emergence. The other example is the origination of polar cyclones in the winter period. We recall that the area of generation of these cyclones coincides with the polar cap region of the terrestrial magnetic field. The onslaught of the solar wind plasma causes the stretching of the magnetic force lines of the polar cap into the tail of the magnetosphere. As a result, the region of the polar cap becomes open to the injection of solar and galactic cosmic radiation into the geomagnetic field with subsequent permeation into the atmosphere. During the re-connection of the magnetic field lines in the Earth's magnetotail, the betatron mechanism causes the speed-up of charged particles. The ionization of atmospheric constituents, including due to the penetration of cosmic radiation up to the tropospheric altitudes, plus the phase transitions of atmospheric humidity lead to the origin of the cyclonic type cells in the atmosphere, where the distribution of aerosol impurity possesses some nonhomogeneous mosaic structure.

Aerosol impurity has a noticeable impact on the global and regional weather and climate (Ginzburg et al. 2009). An aerosol particle has some common external electron shell and a low ionization potential in contrast to potentials for individual atoms in its composition. The background level of the concentration of ionized aerosol in the planetary atmosphere is connected with the charging particles due to friction. Aerosols are also charged by flows of photons, as well as beams of cosmic rays of solar and galactic origination. In addition, the aerosol ionization can be connected with anthropogenic processes. Volcanic eruptions supply a large amount of smog particles. Lightnings are frequently manifested therewith. Therefore, in this process, plasma is also ejected up to the atmosphere. A method of ionization can be thermal in such a case, analogous to the frictional mechanism. During such explosions the kinetic energy of the particles motion produces ionization (some energy transformation). The authors (Bondur, Pulinets 2012) indicate that aerosol particles can make a significant impact on the generation of vortices in the Earth's atmosphere. (Fan et al. 2018) point out that small aerosol particles (with the diameter less than 50 nanometers) considerably speed-up air convection and precipitation. Cyclonic type formations concentrate the energy of latent heat, which is extracted with the help of aerosol components due to phase transitions when moisture condensation and crystallization occurs. As it is demonstrated in (Izhovkina 2014; Izhovkina et al. 2016), the mosaic cellular distribution of charged aerosol subsystems in the presence of a planetary magnetic field brings to the emergence of plasma vortices on pressure gradients orthogonal to the field lines of force. The generated plasma vortices and the large-scale Rossby vortices interact between themselves as particle velocity vortices. The emergence of plasma cellular structures is also manifested in electrical discharges, namely the so-called beaded lightning. This is a rather rare, but rather distinct phenomenon. Characteristics of development of atmospheric vortex formations are affected by electric fields emerging in unsteady aerosol plasma.

Such powerful vortex structures, as tornadoes and typhoons, emerge at low latitudes. The typhoon possesses the shape of an eye. Numerous flashes of lightning occur in the wall of the eye. Using the global data on hurricanes and the data of remote sensing of lightning flashes, (Leary and Ritchie 2009; Price et al. 2009; Fierro et al. 2011) pointed out the following regularity: reinforcement of the tropical cyclone often occurs after an growing of lightning activity in the wall of the eye. Therefore, plasma vortices can have a noticeable effect on the development of powerful vortex formations. The complex vortex structure of a typhoon is visible not only in the picture of lightning discharges. Plasma vortices in the structure of a typhoon are presented in the form of funnels hanging to the earth surface. Typhoon represents, inter alia, some analog of a complex cellular electro-magneto-hydrodynamic generator (Artekha, Belyan 2013; Izhovkina et al. 2016, 2018).

Electrostatic disturbances in nonhomogeneous charged subsystems can significantly effect on the development of plasma vortices. Besides, the presence of plasma vortices gives additional steadiness to large-scale air vortex structures. The energy of the electric field accumulated in clouds is released in lightning discharges when the electric field reaches the breakdown value of the atmospheric layer. There exist linear, horizontal, vertical, ribbon, branching, beaded, volcanic and ball lightning, as well as discharges on high-voltage lines. It is of interest to study the electrostatic unsteadiness of plasma inhomogeneities, since the development of plasma vortices is connected with the emergence of a mosaic cellular structure of the charged aerosol distribution.

One of aims of this article is to show that aperiodic electrostatic fluctuations of nonhomogeneous aerosol plasma can take a remarkable part in the development of air vortex structures. In fact, the inhomogeneity of charged subsystems represents some nonmonotonic filter with respect to electrostatic disturbances. Aperiodic electric fields are generated in plasma irregularities. For the direction along the magnetic field, the Lorentz force equals to zero. In this case, the non-magnetized plasma approximation can be used. In the work, the calculations were performed in the approximation of Boltzmann kinetic equation for hot and cold plasma. The dielectric constant of electrostatic disturbances of inhomogeneous plasma depends non-monotonously on the coordinates. This leads to nonmonotonic separation of plasma inhomogeneities and to the appearance of a mosaic cellular structure. In such structures, plasma vortices are excited upon uneven heating. Lightning discharges when the electric field reaches the breakdown value of the atmospheric layer serve as an indicator for areas of powerful nonlinear pumping of the electric field. The chain structures of electrostatic disturbances are visually observed in lightning discharges - the so-called beaded lightning. A description of this phenomenon is another goal of our article.

*2. The effect of electrostatic perturbations on the activity of plasma vortices and the structure of lightning*

As it is well known, the excitation of air vortex formations and oscillations are energetically connected with the presence of a local excess of free energy (kinetic or potential) for some layer of the planetary atmosphere. According to the data of aircraft and balloon measurements, charges and strong electric fields were detected inside the atmospheric clouds. Connected with them plasma vortices can influence on the development of vortex formations in the Earth's atmosphere (Artekha, Belyan 2013; Izhovkina et al. 2016; Sinkevich et al. 2017). Aerosols can also affect atmospheric phenomena. Aerosol ionization enhances the excitation of vortex formations. The electric field of a plasma vortex is generated analogously to the same field of some MHD-generator. In addition, some aperiodic fields also arise in originating plasma irregularities. In doing so, aperiodically amplifying electrostatic perturbations can be already detected in the simplest approximation of cold plasma, which is nonhomogeneous from arbitrary coordinate.

For heterogeneous plasma, its separation into the cellular structures arises by stochastic manner. The emergence and development of cellular formations is connected with the electrostatic unsteadiness of nonhomogeneous plasma. Some component of the generated atmospheric electric field can arise parallel to the lines of force of the Earth's magnetic field. In such a case, a nonmonotonic "beaded-like" structure of electrostatic disturbances can appear in the approximation of hot inhomogeneous plasma. Apparently, some collective process manifests itself in lightning. (Artekha, Belyan 2018) suppose for the arising channel of lightning that metal-like features of some region in a thunderstorm atmosphere are formed in the unified process. Moreover, the properties of this "medium" can be modulated by various influences. Such a modulation along the lightning channel via electrostatic disturbances can lead to the origination of a chain structure manifested in beaded lightning.

We will seek analytical solutions for the dielectric constant of hot and cold inhomogeneous plasma in the kinetic approximation. To find the particle distribution function, we start from the Vlasov equation with self-consistent fields:

$$\left\{ \frac{\partial}{\partial t} + \mathbf{v}\nabla + \frac{e}{m}\left(\mathbf{E} + \frac{\mathbf{v}\times\mathbf{B}}{c}\right)\nabla_{\mathbf{v}} \right\} f = 0.$$

We will consider the non-magnetized plasma. As applied to the Earth's atmosphere, this approach is strictly applicable to the excitation of electrostatic perturbations along the lines of force of the magnetic field and approximately applicable in other cases, when the effect of the Earth's magnetic field is relatively small compared to other influences. In the absence of a magnetic field, by transformations

$$f(\mathbf{v},\mathbf{r},t) = \frac{1}{(2\pi)^3}\int \exp(i\mathbf{K}\mathbf{v})\tilde{f}(\mathbf{K},\mathbf{r},t)\,d\mathbf{K},$$

$$\tilde{f}(\mathbf{K},\mathbf{r},t) = \int f(\mathbf{v},\mathbf{r},t)\exp(-i\mathbf{K}\cdot\mathbf{v})\,d\mathbf{v},$$

we obtain

$$\left(\frac{\partial}{\partial t} + i\nabla\nabla_{\mathbf{K}} + \frac{ie}{m}\mathbf{K}\cdot\mathbf{E}\right)\tilde{f} = 0.$$

Macroscopic quantities, such as density, flux, energy density, can be expressed through the particle distribution function in a simple manner:

$$n(\mathbf{r},t) = \int f(\mathbf{v},\mathbf{r},t)\,\mathrm{d}\mathbf{v} = \tilde{f}(\mathbf{K}=0,\mathbf{r},t),$$

$$\Gamma(\mathbf{r},t) = \int \mathbf{v}f(\mathbf{v},\mathbf{r},t)\,\mathrm{d}\mathbf{v} = i\nabla_{\mathbf{K}}\tilde{f}(\mathbf{K},\mathbf{r},t)\big|_{\mathbf{K}=0},$$

$$\mathbf{P}(\mathbf{r},t) = m\int \mathbf{v}\mathbf{v}f(\mathbf{v},\mathbf{r},t)\,\mathrm{d}\mathbf{v} = m\nabla_{\mathbf{K}}\nabla_{\mathbf{K}}\tilde{f}(\mathbf{K},\mathbf{r},t)\big|_{\mathbf{K}=0},$$

$$P(\mathbf{r},t) = \frac{m}{2}\int \mathrm{v}^2 f(\mathbf{v},\mathbf{r},t)\,\mathrm{d}\mathbf{v} = \frac{1}{2}\mathrm{Trace}(\mathbf{P}).$$

Substituting $f \to f_0 + f$, in the linear approximation for $\mathbf{E} \sim f$, we get:

$$\left(\frac{\partial}{\partial t} + i\nabla\nabla_{\mathbf{K}}\right)\tilde{f} = -\frac{ie}{m}\mathbf{K}\cdot\mathbf{E}\tilde{f}_0.$$

The Fourier analysis, including the convolution theorem and Parseval's identity, can be used for the problem under consideration (Arfken, Weber, 2001; Bracewell, 1999). The above equation can be solved using an integrating factor. In the case of small initial perturbations *f*, the solution has the form:

$$f(t) = \int_0^\infty \exp(-i\nabla\nabla_{\mathbf{K}}\tau) F(t-\tau)\,\mathrm{d}\tau,$$

where

$$F(t-\tau) = -\frac{ie}{m}\mathbf{K}\cdot\mathbf{E}\tilde{f}_0(\mathbf{K},\mathbf{r},t-\tau).$$

To calculate the dielectric constant, we use the substitution of the potential $\mathbf{E} = -\nabla\varphi$ for the differential form of Gauss's law $\mathrm{div}\,\mathbf{E} = 4\pi e\int f\,\mathrm{d}\mathbf{v}$ and the Poisson equation $\Delta\varphi = -4\pi e\int f\,\mathrm{d}\mathbf{v}$ for electrostatic disturbances *f*. As a result, we obtain the dispersion equation in the operator $(\mathbf{A} \equiv -i\tau\nabla)$ form:

$$\int k^2 \tilde{\varphi}(\mathbf{k},\omega) \exp(i\mathbf{k}\cdot\mathbf{r}) \, d\mathbf{k} = -\frac{4\pi e^2}{m} \times$$

$$\left\{ \int_0^\infty \exp(i\omega\tau) \, d\tau \times \exp(\mathbf{A}\nabla_\mathbf{K}) \int \mathbf{k} \exp(i\mathbf{k}\mathbf{r}) \tilde{\varphi}(\mathbf{k},\omega) \mathbf{K} \tilde{f}_0(\mathbf{r},\mathbf{K}) \, d\mathbf{k} \right\}_{\mathbf{K}=0}. \quad (1)$$

So far, we have not made any assumptions regarding the particle distribution function.

Fourier solutions for collisionless Vlasov-Poisson plasmas are discussed in (Mouhot, Villani, 2011) in a lucid and comprehensive manner.

In the hot plasma approximation, we define the distribution function of particles (e.g., electrons) one-dimensional and inhomogeneous along the $z$ axis in the form

$$f_0 = (2\pi)^{-1/2} N_0 \exp(-z^2/b^2) \alpha^{-1} \exp(-v_z^2/\alpha^2),$$

where $\alpha$ is the thermal velocity for the Maxwell distribution of particles in velocity. Taking into account the orthogonality of the spectral components from (1), we obtain the dispersion equation of electrostatic perturbations:

$$\varepsilon = 1 + \frac{\sqrt{\pi} b \omega_p^2}{\sqrt{2} k^2 \alpha^2} \exp(-ikz) \times$$

$$\int_{-\infty}^{\infty} dk_2 \left\{ \frac{k}{k_2} 2(xZ(x)+1) \exp[-(k_2-k)^2 b^2/4] \exp(ik_2 z) \right\} = 0, \quad (2)$$

where $Z(x)$ is the plasma dispersion function, $x = \omega/(k_2 \alpha)$,

$$Z(x) = i \int_0^\infty d\tau \exp(ix\tau - \tau^2/4).$$

It can be seen from the expression (2) for $\varepsilon$ that the plasma inhomogeneity is a nonmonotonic filter with respect to electrostatic disturbances.

For a cold plasma with a symbolic delta function in particle velocity

$$f_0(z, v_z) = N_0 \exp(st) \exp(-z^2/b^2) \delta(v_z)$$

for $s < 0$, the dispersion equation of electrostatic perturbations of the plasma inhomogeneity has the form

$$\varepsilon = 1 + \omega_p^2 \exp(st)\exp(-z^2/b^2) \times$$

$$\frac{[1+2zi/(kb^2)](s^2 - \omega^2 + 2i\omega s)}{(s^2 + \omega^2)^2} = 0. \qquad (3)$$

Upon transition to a homogeneous plasma $s \to 0$, $b \to \infty$, from (3) we obtain $\varepsilon = 1 - \omega_p^2/\omega^2 = 0$, that is, the dispersion equation of a cold homogeneous plasma, where $\omega_p^2 = 4\pi N_0 e^2/m$ is the square of the plasma frequency.

From (3), a tendency is seen – the appearance of an aperiodic electrostatic disturbance in a plasma inhomogeneity at frequencies below the local plasma. For

$$f_0(z, v_z) = N_0 \exp(-z^2/b^2)\delta(v_z)$$

we have:

$$k(\omega, z) = \frac{(2zi\omega_p^2/b^2\omega^2)\exp(-z^2/b^2)}{1 - (\omega_p^2/\omega^2)\exp(-z^2/b^2)} .$$

As characteristic Vlasov-Poisson singularity the dispersion relation features a dipolar border line. Here, the magnitude of the wave vector $k$ determines the spatial scale of electrostatic disturbances.

Chain structures in unstable nonhomogeneous plasmas also appear in lightning discharges. As the temperature and pressure in the lightning channels increases to enormous values, the influence of the geomagnetic field on the plasma weakens – lightning sparks in arbitrary directions. The nonmonotonic "beaded" structure of electrostatic perturbations of a hot inhomogeneous plasma also manifests itself according to the results of analytical calculations of the dielectric constant: the factor exp(–ikz) in front of the integral in formula (2) modulates the structure of the partially ionized aerosol subsystem (the influence of a similar term under the integral is smoothed by integration). The instabilities of hot nonhomogeneous plasma were analyzed in the kinetic approximation. It follows from the analysis performed that electrostatic perturbations of nonhomogeneous plasma provoke the formation of cellular structures.

Plasma inhomogeneity represents some nonmonotonic filter with respect to electrostatic disturbances. The nonmonotonic alternate fibering of the inhomogeneity promotes the formation of a mosaic structure and the excitation of plasma vortices in such a structure. In the lightning channel, the nonmonotonic alternate stratification of electrostatic disturbances of nonhomogeneous plasma is observed visually in the form of beaded lightning. One of existing opinions that channel can look as beaded (or chain) only for concrete observer due to the different direction of a discharge and the preferred relative location of this observer, is erroneous. The matter is that discharges occur in many different directions during any thunderstorm. Besides, one and the same lightning discharge is often fixed by many observes from various positions. However, despite of all these observations, the phenomenon of beaded (or bead, chain) lightning is few and far between. Disagreements in the results of observing the same lightning have also not yet been recorded. So, the explanation with using variations in directions, illuminations (seeming brightness) or overlapping the channel by rain clouds (shadowiness) is faulty. Farther, as a consequence of the incomparability in order of magnitude for the rate of development of the electric discharge and the characteristic speed of air movement, the explanation of the occurrence of the beaded lightning via air cooling the lightning channel seems very doubtful. Besides, characteristics of the medium inside and outside the lightning channel differ dramatically and the rapid cooling of separate segments due to difference in thickness (radii) of the lightning channel is also not realistic. Therefore, the chain structure should be formed immediately together with the preparation and the emergence of the lightning channel. This assumption is also confirmed by the fact that the duration of a beaded lightning is larger than that for an usual lightning flash. The several longer duration of a beaded lightning as compared to a conventional lightning is understandable, since only part of the path corresponds to breakdowns, and the remaining sections resemble a dark current. As a result, the total time (averaged) of the discharge process is increased. Moreover, the visible structure of beaded lightning may be rather orderly, wave-like. Thus, from the viewpoint of the mechanism of the phenomenon, the modulation of the lightning channel medium through electrostatic disturbances is a fairly reasonable explanation for the occurrence of beaded lightning.

What is the characteristic scale of such disturbances ($\bar{k} = 2\pi/\bar{\lambda}$) that could be visualized? Modulations too small in $\bar{\lambda}$ will have nothing effect on the electrical breakdown (the lightning channel simply will not "notice" them, as well as the external observer). It would be difficult for too large modulations in $\bar{\lambda}$ to achieve noticeable values due to the presence of stratification and variable heterogeneity of the cloud masses; in addition, the effect of large-scale modulations on lightning would most likely come down to curvatures of the trajectory of the lightning discharge (since the current is trying to search for the "easiest ways"). As a result, only modulations of intermediate scales can lead to observable chain structures.

If we proceed from the fact that the recorded inhomogeneities of the phenomenon under consideration are meters — several tens of meters, then we can very roughly estimate the probability of occurrence of beaded lightning among ordinary lightnings as the value less than the ratio of such characteristic scales ($\bar{\lambda}$) to the storm cloud scale. Considering additionally that such perturbations still have to have time to grow up to noticeable values (and to possess some steadiness), the probability of realization of conditions for the appearance of beaded lightning among ordinary lightnings is much less than 0.001.

If the situation with the only predominant scale of electrostatic modulation $\lambda_1$ is realized, then the chain structure will be rather regular (ordered). But if there arose several disturbances of different scales $\lambda_i$, comparable in amplitude, then the structure of beaded lightning can be more complex (seemingly disordered).

Thus, modulation by electrostatic disturbances is the most likely cause of the occurrence of beaded lightning. Moreover, such chain structure can contribute to the excitation or modulation of other types of waves (electromagnetic, acoustic, etc.).

The ordered beaded structure is especially reminiscent of standing waves in a waveguide (when half the wavelength is stacked in the transverse direction). The temperature in the lightning channel itself rises up to $10^4$ degrees, the pressure reaches hundreds of atmospheres, the current strength reaches hundreds of thousands of amperes and above; the main channel of lightning has therewith a diameter of 10 to 25 cm. However, if there was already a lightning strike nearby (i.e. this is a repeated discharge), then the environmental properties changed around that lightning channel at a distance of meters (sometimes tens of meters), which can constitute a natural waveguide for the development of subsequent beaded lightning. But this is only one of the possible mechanisms for preserving perturbations of the required spatial scale.

For ionized aerosol subsystem in the atmosphere, the cold plasma approximation is applicable except for areas of spark discharges, where strong electric fields in the atmosphere become apparent. For lightning channels, it is necessary to consider the approximation of hot inhomogeneous plasma. Such calculations of the parameters of electrostatic instability are presented above for both cases. In cold inhomogeneous plasma, aperiodic electrostatic deviations spring up along the geomagnetic field, and, as a result, mosaic cellular structures of electric fields and charge density are shaped. Upon uneven heating of such the cellular structures, atmospheric plasma vortices are excited.

A remarkable part in the distribution and dynamics of atmospheric plasma nonhomogeneous subsystems is taken by rising electric fields and originating plasma vortices, which enhance the activity of air vortices in the planetary atmosphere (Artekha, Belyan 2013; Izhovkina et al. 2016; Sinkevich et al. 2017). The excitation of plasma vortices in nonhomogeneous mosaic distributions of ionized subsystems upon moisture condensation on aerosols and inhomogeneous heating of such structures affects the origination and development of large-scale vortices – of cyclonic and anticyclonic types.

Electric fields give mobility to plasma vortices, accelerate their interaction. In a medium with double gyrotropy, atmospheric vortex formations amplify, since plasma vortices and Rossby vortices cooperate as particle velocity vortices. An increase in the power of vortex structures – tornadoes, hurricanes, and anticyclones – is facilitated by the growth of the concentration of atmospheric contamination taking into account aerosol impurities.

The elasticity of the vortex structures is given, inter alia, by electric fields, currents and plasma vortices. For example, in the interrelation of a cyclone and a non-blocking anticyclone, one structure is pushed by another. As a result, there arises the joint drift of the cyclone – anticyclone pair. The interaction between them differs from a simple mixing of air masses. Plasma vortices provide some additional sustainability to atmospheric vortex structures. Electromagnetic interactions between plasma vortices affect the trajectories of the cyclone and anticyclone as they approach each other. So, the picture of the interaction of the cyclone – anticyclone pair during the accumulation of contaminants, their ionization by external sources, and the strengthening of the anticyclone structure under the influence of plasma rotary motions in a geomagnetic field changes significantly. For example, a blocking anticyclone deflects the movement of the cyclone approaching from the west, from the Atlantic Ocean, to the north, while the anticyclone is stationary and continues to grow. The growth of the anticyclone is caused by the heating of the underlying surface and atmosphere by solar radiation, the conglomeration of contaminants, their ionization and the impact of aerosol plasma on the vortex activity.

## 3. *Conclusion*

When mosaic cellular plasma formations are heated at pressure gradients orthogonal to the geomagnetic field, an atmospheric electro-magnetohydrodynamic generator with plasma vortices in the structure cells begins to act. As some process of the reaction and integration of plasma vortices between themselves and of the cooperation between plasma vortices and Rossby vortices, atmospheric vortex structures of tornadoes, cyclones, and anticyclones are formed and grown. Part of the energy of such the structures is therewith pumped by plasma formations.

The process of developing aerosol plasma instabilities can be seen in the manifestations of atmospheric electricity. Cellular structures are also formed in lightning discharges. In sometimes appearing beaded lightning, the distinct chain structures are observed visually. From the analytical derivations of the instability parameters of hot inhomogeneous plasma, the formation of such structures seems to be stochastically predestined. Electrostatic perturbations of inhomogeneous plasma with the Earth's magnetic field are connected with the electro-magnetohydrodynamic effect, namely, the generation of electric fields in plasma flows with pressure gradients, which are orthogonal to some external magnetic field.

The article presents analytical calculations of the instability parameters for hot inhomogeneous plasma without a magnetic field. As applied to the geophysics, such solutions can be attributed to the direction of electric fields coincided with the geomagnetic field direction, as well as for areas of hot plasma in lightning flashes (when the impact of the Earth's magnetic field is relatively small). The kinetic approximation was used in the presented theoretical analysis, where the distribution of particles in the velocity space is taken into consideration. In the cold plasma approximation, such a distribution is determined by the delta function. The nonmonotonic dependence of the dielectric constant of electrostatic disturbances on the coordinates leads to the appearance of a cellular inhomogeneous plasma structure in electric fields excited in a plasma inhomogeneity at plasma density gradients. The projection of electric fields on the direction of geomagnetic field lines accelerates the integration process for plasma vortices in a geomagnetic field tube that can enhance the large-scale vortex structure as a whole.